# Systematic strain-induced bandgap tuning in binary III–V semiconductors from density functional theory

Badal Mondal[1,2], Ralf Tonner-Zech[1*]

[1] Wilhelm-Ostwald-Institut für Physikalische und Theoretische Chemie, Universität Leipzig, 04103 Leipzig, Germany
[2] Fachbereich Physik, Philipps-Universität Marburg, 35032 Marburg, Germany

E-mail: ralf.tonner@uni-leipzig.de

**Abstract**

The modification of the nature and size of bandgaps for III-V semiconductors is of strong interest for optoelectronic applications. Strain can be used to systematically tune the bandgap over a wide range of values and induce indirect-to-direct (IDT), direct-to-indirect (DIT), and other changes in bandgap nature. Here, we establish a predictive *ab initio* approach, based on density functional theory, to analyze the effect of uniaxial, biaxial, and isotropic strain on the bandgap. We show that systematic variation is possible. For GaAs, DITs were observed at 1.52% isotropic compressive strain and 3.52% tensile strain, while for GaP an IDT was found at 2.63 isotropic tensile strain. We additionaly propose a strategy for the realization of direct-indirect transition by combining biaxial strain with uniaxial strain. Further transition points were identified for strained GaSb, InP, InAs, and InSb and compared to the elemental semiconductor silicon. Our analyses thus provide a systematic and predictive approach to strain-induced bandgap tuning in binary III-V semiconductors.

## 1. Introduction

Semiconductor compounds attract a great amount of attention, both in science and technology, due to their immense application range in areas such as optoelectronics and integrated circuits [1,2]. One of the major goals in basic and applied research is to tailor the optical properties of semiconductor materials to a target application. The most important fundamental property determining these properties is the material's bandgap. For example, materials for optical telecommunication applications require direct bandgaps in the range of 0.80–0.95 eV [3–5], while a range of 0.5–2.0 eV is necessary for materials used in efficient solar cells [6–9]. One material class that is especially versatile in this respect are compound semiconductors, specifically the III-V semiconductors composed of elements from group 13 and 15 of the periodic table of elements [2,5,10–22]. In the last decades, the optical properties of this material class have been intensively investigated [2,10–14,22–36] and several strategies have emerged to fine-tune the bandgap. Changing the nature of the chemical elements and their relative composition is a powerful approach to vary the gap over a wide range of energies [37–44]. However, changing the chemical composition is not always possible. One reason for this lies in the constraints in the growth characteristics of precursor molecules for the chemical vapor deposition techniques often used to grow these materials. Another reason is the thermodynamic instabilities of some elemental

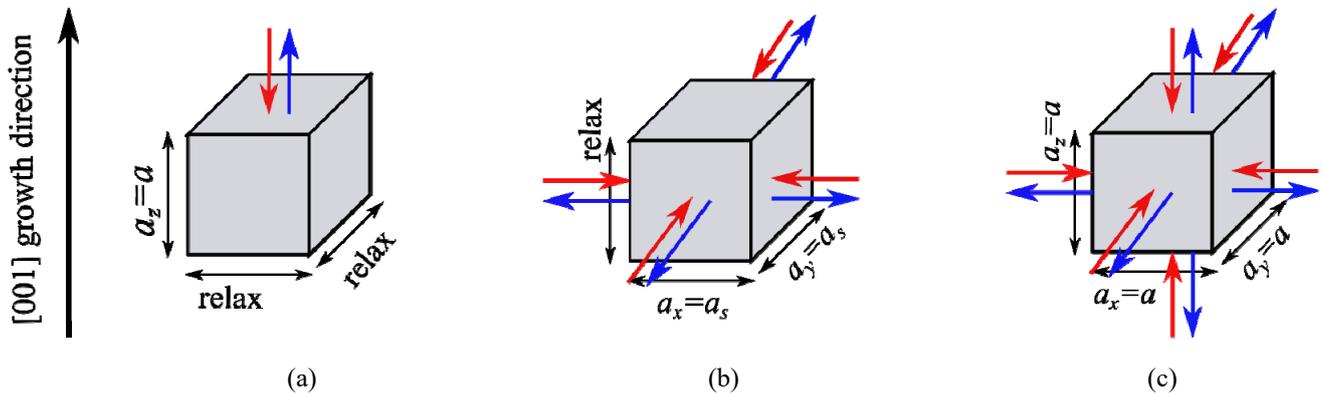

Figure 1: The strain models used in this study: (a) uniaxial strain, (b) biaxial strain, and (c) isotropic strain. The blue and red arrows correspond to tensile and compressive strain, respectively. The *z*-direction is defined as the strain (a) or growth (b) direction. For the biaxial case, $a_s$ corresponds to the (epitaxial) substrate lattice parameter.

compositions [3,14,37,44–48].

An alternative and sometimes also complementary approach to vary the bandgap is strain engineering. This can be achieved through external effects such as: applying pressure on the system [23,24,27–36], altering the temperature of the system, or changing the substrate in epitaxial growth processes [13,16,19,37–42,45,46,49–62]. All of these approaches result in structural strain in the system because of the deviation of one or several lattice parameters of the material from their equilibrium values. The effects on the electronic structure from straining the material are shifts in valence-band maximum (VBM) and conduction-band minimum (CBM), and thus, the variation in the bandgap.

Many attempts have been made to understand the effect of strain from theoretical perspectives. The electronic properties of semiconductor materials have been previously analyzed by: (a) empirical or semi-empirical methods such as the local/nonlocal empirical pseudopotential method [63–71], the semi-empirical tight-binding method [72–84], the k·p method [85–90]; or by (b) *ab initio* methods [91–98] such as density functional theory (DFT) [91,99–104]. Although empirical and semi-empirical methods are computationally efficient and often easy to apply, they rely on many empirical fitting parameters. This strongly lowers their ability to predict properties for new materials, which is a core goal in computational materials design [105]. In contrast, *ab initio* methods allow the calculation of the electronic structure without the need for empirical fitting parameters.

One of the most widely used *ab initio* approaches in material science is DFT. The crucial ingredient here is the density functional. Functionals following the generalized gradient approximation (GGA) often lead to an excellent agreement of computed lattice parameters with experimental data. However, they are known to show very large errors for bandgaps [105–109]. Hybrid functionals such as HSE06 [110] and GW-based methods [95–98] can solve this issue but are computationally expensive. Previously, we and others successfully used the exchange-correlation functional, developed by Tran and Blaha, to predict the electronic properties of unstrained III-V compound semiconductors without empirical adjustments or application of scissor operators [105,111–117]. Here, we will show that this approach can also successfully be used for predicting properties by applying a wide range of strains to these materials.

Although strain engineering is an established field for IIIV semiconductors [2,45,87,118,119], the investigation of strain effects has recently found renewed interest in the field of nanowires [120–128]. Furthermore, no systematic theoretical study is yet available that predicts optical properties of strained materials of this kind without empirical adjustments. We now set out to reliably predict the optical properties of strained materials over a wide range of strains. This will ultimately enable computational materials design approaches in strain engineering of established and upcoming materials.

This work will establish the methodology and highlight the challenges of predictive modelling. Thus, we focus our analysis of strain effects on the electronic structure of the most widely investigated binary III-V semiconductors: GaAs and GaP. These materials are not only interesting for basic research but also support a wide range of applications (either as binary materials or as a host material for multinary compounds) in microelectronics, solar cells, laser technology, and LEDs [2–9,14–21]. To show the general applicability of our approach, selected data on the materials Si, GaSb, InP, InAs, and InSb are included. The ultimate goal is to provide guidelines for future experimental work on strained materials.

**2. Model**

Table 1: Computed (PBE-D3(BJ)) equilibrium lattice parameters (Å) for the materials investigated in comparison to experimental reference values.

| System | Si | GaP | GaAs | GaSb | InP | InAs | InSb |
|---|---|---|---|---|---|---|---|
| Calculation | 5.421 | 5.474 | 5.689 | 6.134 | 5.939 | 6.138 | 6.556 |
| Experiment [38,129] | 5.431 | 5.451 | 5.653 | 6.096 | 5.869 | 6.058 | 6.479 |

In this study we model uniaxial, biaxial, and isotropic strain. Figure 1 schematically shows which lattice parameters are kept fixed and which are relaxed in the modelling of these three types of strain. The materials investigated all feature zincblende structures. The growth direction in the "theoretical epitaxy" approach applied (see below) was taken to be [001] in this study and was defined as the *z*-direction in our modelling approach. We limited our analysis of uniaxial strain to the application of strain along the growth direction only. This uniaxial (compressive) strain is experimentally realized most often by applying pressure. Here, we modelled this by varying the lattice parameter in the *z*-direction while relaxing the in-plane lattice parameters (figure 1a). In the spirit of a systematic study, we also studied uniaxial tensile strain over the same range of values. However, the experimental realization of expanding lattice parameters is typically limited to small strain values (e.g. by increasing the temperature or applying shear stress). The major approach to produce biaxially strained materials is epitaxial growth on a substrate with a different lattice parameter ($a_s$). We thus fixed the in-plane lattice parameters ($a_x$, $a_y$) to the lattice parameter of an (imaginary) substrate ($a_s$) while varying the parameter in the growth direction (figure 1b). In this case, we considered the structural strain imposed by a substrate but neglected the electronic influence for the modelling (theoretical epitaxy) [45,119]. Isotropic strain is then consequently modelled by not constraining any lattice parameter and increasing (decreasing) all lattice parameters by the same amount (figure 1c).

## 3. Computational details

Computations were carried out with DFT-based approaches as implemented in the Vienna Ab initio Simulation Package (VASP 5.4.4) [101,130–133], using plane-wave basis sets in conjunction with the projector augmented wave (PAW) method [134,135]. The primitive zinc blende cell was used throughout. The basis set energy cutoff of 450 eV, the electronic energy convergence criteria of $10^{-6}$ eV, and the force convergence of $10^{-2}$ eV/Å were used. Reciprocal space was sampled with a 10×10×10 Γ-centered Monkhorst-Pack k-point mesh [136]. Optimizations of the primitive cells were performed using the PBE functional [106] with the semi-empirical dispersion correction scheme DFT-D3 with a Becke-Johnson type damping function [137,138]. The geometry optimizations were carried out by the consecutive volume and position optimization until convergence was reached. For every set of lattice parameter values investigated, all atomic positions were optimized.

For the bandgap and band structure calculations, the TB09 functional was used [111] including spin-orbit coupling. This had previously been used to give an excellent agreement with the experimental bandgaps for this compound class [105,111–116]. The band energies for all the different configurations were re-normalized to the respective VBM. We limited our calculations to a range of ±10% strain by applying constrained optimizations as outlined in the previous section. This is in the order of magnitude of pressures (10 to several 100 GPa) achievable in modern experiments [33,139–150]. We indicate tensile strain with a "+" sign to emphasize the positive strain value and to make it easier for the reader to distinguish it from compressive strain values, which are denoted with a "−" sign.

The contribution of the atomic orbitals at the different k-points on the bands was calculated by projecting the plane waves on the minimal basis set using LOBSTER [151,152].

## 4. Results

### 4.1 Unstrained structures

Before we discuss the influence of strain, the unstrained materials investigated here shall briefly be presented. Regarding the sign convention, we define positive strain to correspond to expansion (tensile strain) and negative strain as compression (compressive strain). The strain values were calculated according to equation 1.

$$\text{strain (\%)} = (a_f - a_{eqm})/a_{eqm} \times 100 \quad (1)$$

Here, $a_f$ is the lattice parameter in the strained structure while $a_{eqm}$ is the equilibrium lattice parameter. The equilibrium lattice parameters for all materials investigated were computed with the PBE-D3(BJ) approach and are given in table 1. The good agreement with the experimental lattice parameters (maximum deviation of 0.09 Å or 1.3%) lends confidence to the accuracy of the theoretical approach.

*4.2 Gallium arsenide*

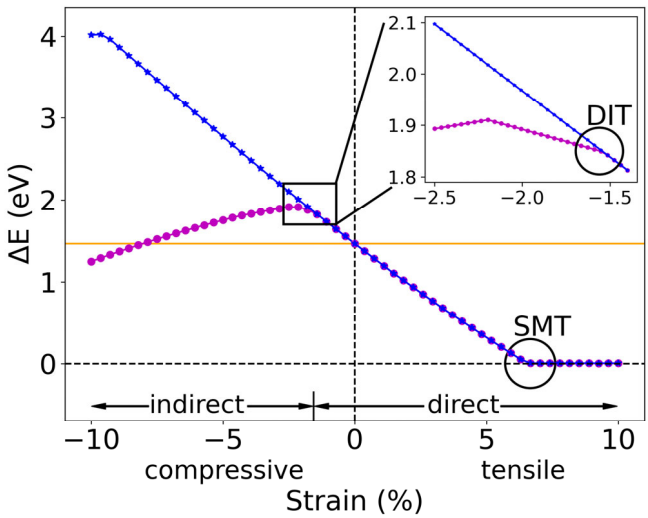

Figure 2: Isotropic strain effects on the bandgap of GaAs. The energy difference between CB and VB at the $\Gamma$-point ($\Delta E_\Gamma$, blue), the bandgap ($E_g$, magenta), the direct to indirect transition (DIT), and the semiconductor to metal transition (SMT) are shown. The solid orange line indicates $E_g$ for the equilibrium structure (1.47 eV). The inset shows the region of the DIT with a finer grid of strain calculations.

*4.2.1 Isotropic strain* First, we present the bandgap variation of GaAs under the application of isotropic strain in a range of ±10% around the equilibrium lattice parameter. Figure 2 shows the variation of the energy difference between the conduction band (CB) and the valence band (VB) at the $\Gamma$ point ($\Delta E_\Gamma$), as well as the bandgap ($E_g$), as a function of strain. Here, and throughout the manuscript, we distinguish between these two energy differences. For a direct bandgap material, both values are the same. If $E_g$ is smaller than $\Delta E_\Gamma$ this indicates an indirect bandgap. A strain value where the two curves start deviating thus indicates a direct to indirect (DIT) bandgap transition point.

Under tensile strain, $E_g$ decreases until, at +6.78% strain, the bandgap vanishes, corresponding to a semiconductor to metal transition (SMT). In this case, the $E_g$ curve coincides with the $\Delta E_\Gamma$ curve throughout, indicating a direct bandgap. Under compressive strain, however, the $E_g$ curve initially follows the $\Delta E_\Gamma$ curve until −1.56% strain, where the $E_g$ curve then separates from the $\Delta E_\Gamma$ curve. Although $\Delta E_\Gamma$ still increases under further strain, the bandgap starts to decrease. Thus, we have a DIT point here.

To understand the origin of this deviation we looked at the change in CB energies for strained GaAs relative to their values for the equilibrium structure at four high-symmetry points in reciprocal space: $\Gamma$, L, $\Delta_m$, and X points (figure 3). This showed that the CB energies at the $\Gamma$ and L points decrease under tensile strain with a slight increase for $\Delta_m$ and X points. The largest change was found at the $\Gamma$ point, followed by L, $\Delta_m$, and X. As unstrained GaAs is a direct bandgap semiconductor, this signifies that the bandgap remains direct under tensile strain, to begin with. Subsequently, at +6.78% strain, the CBM and VBM become degenerate, which results in the SMT.

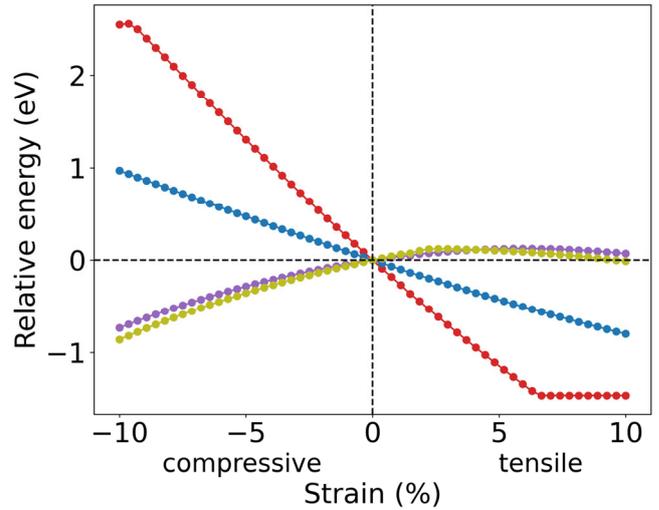

Figure 3: Variation of CB energies for isotropically strained GaAs at the k-points $\Gamma$ (red), L (blue), $\Delta_m$ (purple), and X (olive) relative to their values in the unstrained structure. The band energies at differently strained configurations were re-scaled with respect to their corresponding VBM.

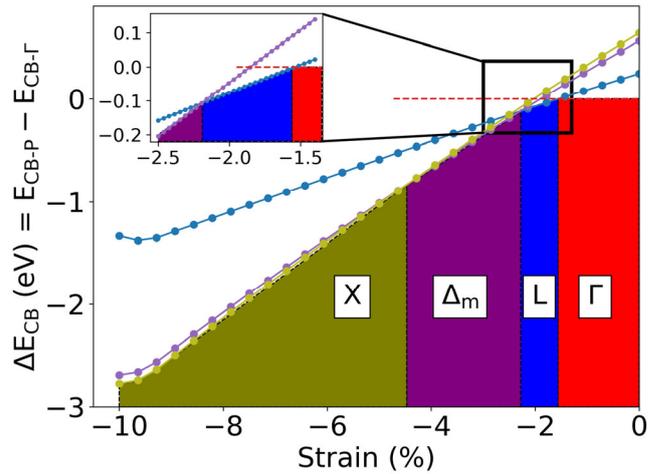

Figure 4: The difference between CB energies at the $\Gamma$ point and other k-points P ($\Delta E_{CB} = E_{CB-P} - E_{CB-\Gamma}$; with P = $\Gamma$, L, $\Delta_m$, and X) for isotropically strained GaAs. Colored areas indicate at which k-point we find the CBM for the given value of compressive strain. The color scheme: $\Gamma$ (red), L (blue), $\Delta_m$ (purple), and X (olive).

For compressive strain, however, the CB energies at the $\Gamma$ and L point increase while a slight decrease was found at the $\Delta_m$ and X points (figure 3). This resulted in an increase in the direct bandgap for small strain values. Since CB at the Γ point changed the most with strain, beyond −1.56% strain it superseded the energy at the L point. This resulted in the CBM shifting from the Γ to the L point and a DIT. CB energy at the Γ point increased further resulting in a steep increase of $\Delta E_\Gamma$ for high strain values (figure 2). CB at the L point increased much slower in energy, flattening the $E_g$ curve and even producing bandgap decrease at high compressive strain (figure 2). The band structures computed at each strain value can be found in the supporting information (figures S1a-b).

Further, by comparing the difference in CB energies at the Γ, L, $\Delta_m$, and X points, different 'transition points' were estimated (figure 4). The first transition was from Γ to L at −1.56% strain (DIT), the second from L to $\Delta_m$ at −2.28% strain, and the third from $\Delta_m$ to X at −4.47% strain. Increasing the resolution in the k points further revealed that, unlike the sharp first and second transitions, the third was much smoother (figure S1c). During the third transition, the CBM started to flatten out with strain, until at −6.78% when the X point also became part of the CBM-plateau. After that, the plateau started to shrink towards the X point. Note, that the VBM remained at the Γ point throughout. Thus, the change in CBM under compressive strain determined the nature of the transitions.

*4.2.2 Biaxial strain* The effect of the biaxial strain on the bandgap is shown in figure 5 (band structures are shown in figures S1d,e). The bandgap decreased under both compressive and tensile strain. During compression, the $E_g$ curve coincided with $\Delta E_\Gamma$ throughout. The CBM always remained at the Γ point, and hence, the bandgap remained direct. Only for very high compressive strain values (beyond −7.86%), did the CBM and VBM become degenerate, leading to an SMT. For tensile strain, a DIT was found at +3.52% strain, exemplified by the $E_g$ curve splitting from $\Delta E_\Gamma$ in figure 5.

Under further tensile strain, the bandgap continued to decrease until GaAs became a semimetal at +8.00% strain. Thus, we observed a semiconductor to semimetal transition (SsMT). By comparing the difference in CB energies at the Γ, L, $\Delta_m$, and X points (figure 6) we found the DIT to correspond to a Γ to $\Delta_m$ transition. No further transition points were found here.

*4.2.3 Uniaxial strain* The uniaxial strain model in our case is equivalent to the biaxial strain model. This is true because we consider the [001] crystal orientation in the zincblende crystal grown on the [001] surface of another zincblende substrate, and the uniaxial strain is then applied in the <100> direction (figure 1a). In this configuration, relaxing the lattice parameter in the z-direction at fixed in-plane (*x* and *y*) lattice parameters ($a_s$) is equivalent to fixing it in the z-direction at the value *a*, and relaxing the in-plane parameters. For the uniaxial strain in other crystal orientations or directions, this equivalence is not true, because of finite off-diagonal stress tensor elements [118,153,154].

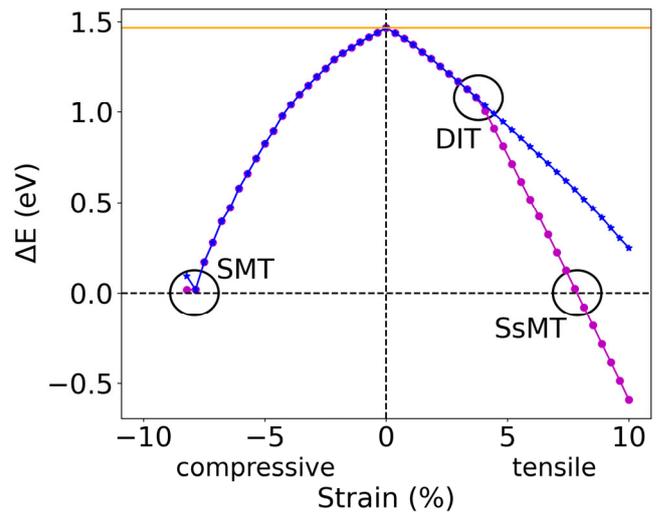

Figure 5: Biaxial strain effects on the bandgap of GaAs. The energy difference between CB and VB at the Γ-point ($\Delta E_\Gamma$, blue), the bandgap ($E_g$, magenta), the direct to indirect (DIT), the semiconductor to metal (SMT), and the semiconductor to semimetal (SsMT) transitions are shown. The solid orange line indicates $E_g$ for the equilibrium structure (1.47 eV).

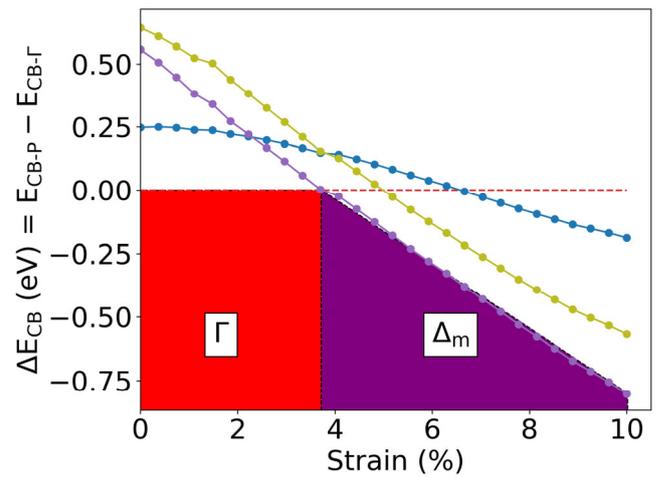

Figure 6: The difference between CB energies at the Γ point and other k-points P ($\Delta E_{CB} = E_{CB-P} − E_{CB-\Gamma}$; with P = Γ, L, $\Delta_m$, and X) for biaxially strained GaAs. Colored areas indicate at which k-point we find the CBM for the given value of tensile strain. The color scheme: Γ (red), L (blue), $\Delta_m$ (purple), and X (olive).

Therefore, we used the data from the previous subsection and now present them as a function of the out-of-plane lattice parameter (figure 7). This is essentially a mirrored version of figure 5 with a changed scaling of the *x*-axis. Now, we found the DIT at −7.30% strain, the SsMT at −15.43% strain, and the SMT at +17.73% strain, respectively.

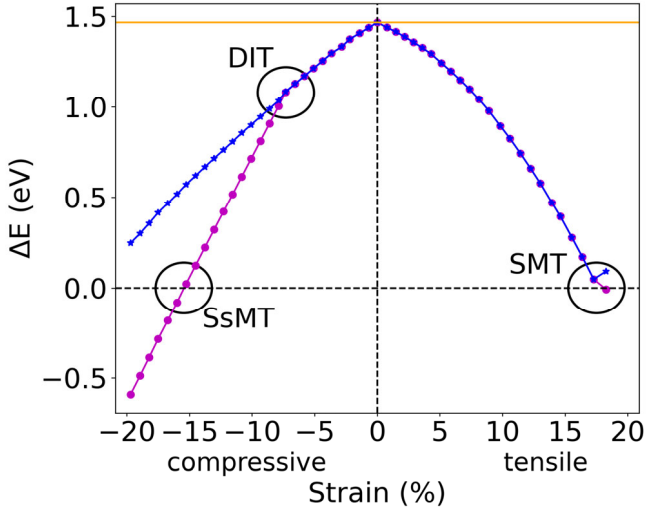

Figure 7: Uniaxial strain effects on the bandgap of GaAs. The energy difference between CB and VB at the $\Gamma$-point ($\Delta E_\Gamma$, blue), the bandgap ($E_g$, magenta), the direct to indirect (DIT), the semiconductor to metal (SMT), and the semiconductor to semimetal (SsMT) transitions are shown. The solid orange line indicates $E_g$ for the equilibrium structure (1.47 eV).

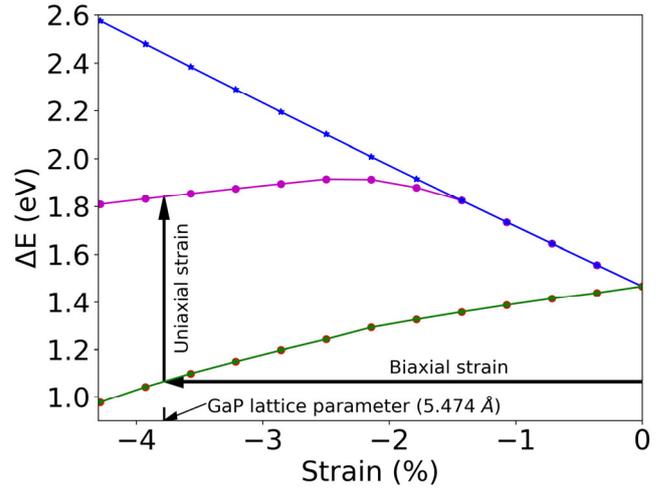

Figure 8: Computational bandgap engineering for GaAs grown on GaP substrate. The energy difference between CB and VB at the $\Gamma$-point ($\Delta E_\Gamma$, blue for iso and green for bi) and the bandgap ($E_g$, magenta for iso and red for bi) are shown as a function of biaxial (bi) and isotropic (iso) strain.

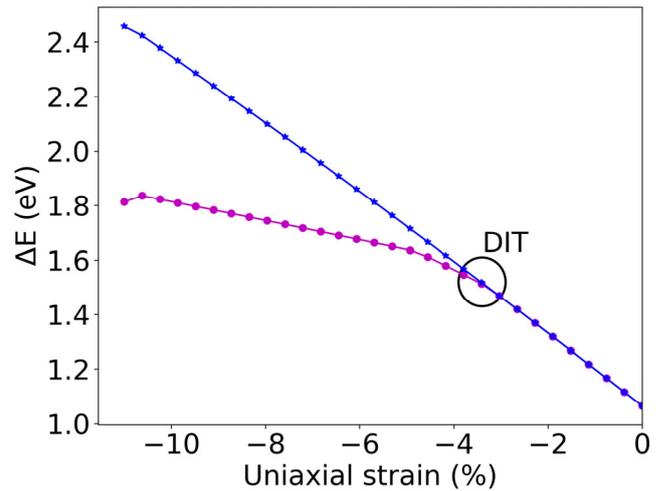

Figure 9: Change in bandgap as a function of uniaxial compressive strain along $z$-lattice parameter for GaAs grown on GaP [001] substrate. The energy difference between CB and VB at the $\Gamma$-point ($\Delta E_\Gamma$, blue), the bandgap ($E_g$, magenta), and the direct to indirect (DIT) transition are shown.

*4.2.4 Combining biaxial and uniaxial strain*   Uniaxial and biaxial strain were shown in the previous sections to be useful strategies to tune the bandgap. However, one major goal in tuning the electronic structure is changing the nature of the bandgap. As shown in the previous subsection, this cannot be achieved via biaxial compressive strain, which is one of the most common experimental realizations of strain via epitaxial growth. We will now show that by combining uniaxial and biaxial strain this can in fact be achieved.

In a thought experiment (figure 8), the desired material (GaAs) is first "grown" epitaxially on a substrate with a smaller lattice constant (e.g., GaP), resulting in compressive biaxial strain (here: −3.78%). Such epitaxial growth would lead to expansion of the $z$-lattice parameter. Subsequently, uniaxial compressive strain (e.g., pressure) could be applied along the $z$-direction to compress the $z$-lattice parameter. We assume that the in-plane lattice parameters would not relax upon compression of the $z$-lattice parameter. This is shown more clearly in figure 9 where we find a DIT point at −3.2% uniaxial strain.

Next, we further generalized this strategy. Figure 10 shows the required uniaxial compressive strain for the DIT in biaxial compressively strained GaAs. No transition can be achieved for biaxial compressive strain below 1.56%, as both the biaxial and isotropic strain would have the same direct nature of the bandgap (figure 8).

A similar strategy can be applied for the indirect to direct transition (IDT) in biaxial tensile strained GaAs. In this case, one would need to expand the $z$-lattice parameter for the transition. Experimentally, this can be achieved e.g., by thermal expansion. As shown in figure 11, this however, is only reasonable for biaxial strain smaller than 4.5%. For higher biaxial strain the large required amount of uniaxial

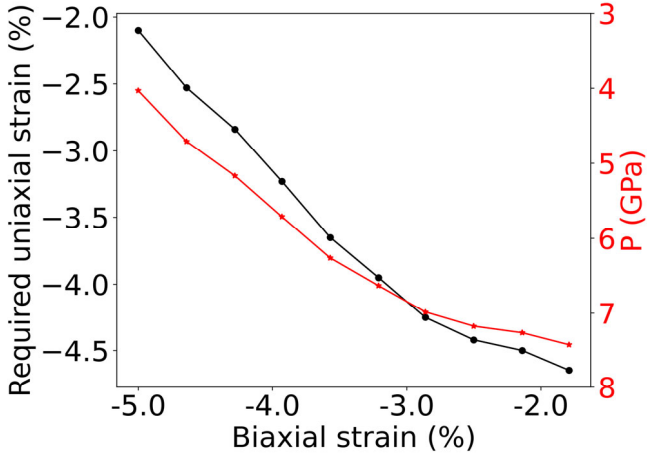

Figure 10: Variation of required uniaxial compressive strain for the direct to indirect transition (DIT) in biaxial compressively strained GaAs. 3rd order Birch-Murnaghan equation [155] was used for the strain to pressure conversion. We used reference [156,157] for the bulk modulus and its first derivative data for the conversion.

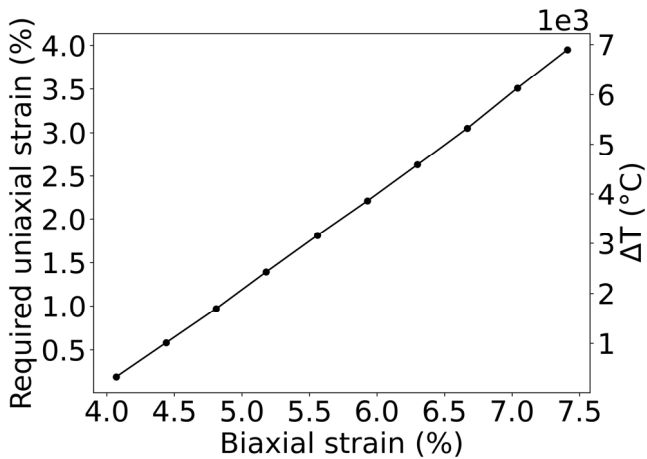

Figure 11: Variation of required uniaxial tensile strain for the indirect to direct transition (IDT) in biaxial tensile strained GaAs. The temperature increase ($\Delta T$) required for the thermal expansion was calculated using the linear thermal expansion coefficient of GaAs ($5.73 \times 10^{-6}$ °C$^{-1}$ [129]).

tensile strain can not be achieved by thermal expansion only.

*4.3 Gallium phosphide*

Gallium phosphide is an indirect bandgap semiconductor. Next, we demonstrate the application of strain on the bandgap engineering. Figure 12 shows the variation in CB energy for GaP under isotropic strain at the $\Gamma$, L, $\Delta_m$, and X points relative to their equilibrium values (band structures are shown in figures S1f,g).

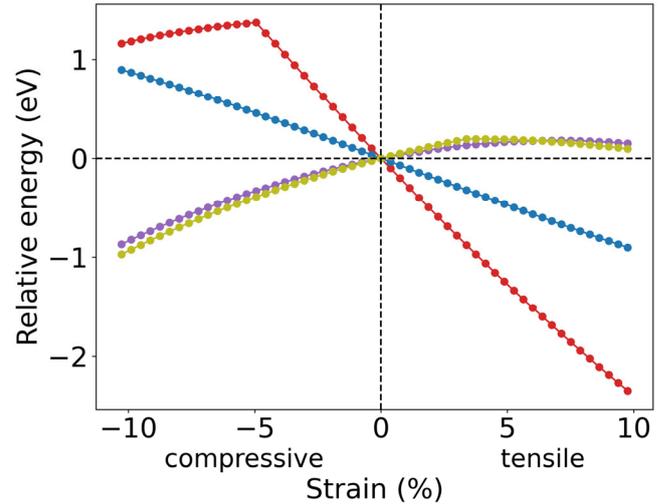

Figure 12: Variation of CB energies for isotropically strained GaP at the k-points $\Gamma$ (red), L (blue), $\Delta_m$ (purple), and X (olive) relative to their values in the unstrained structure. The band energies at differently strained configurations were rescaled with respect to their corresponding VBM.

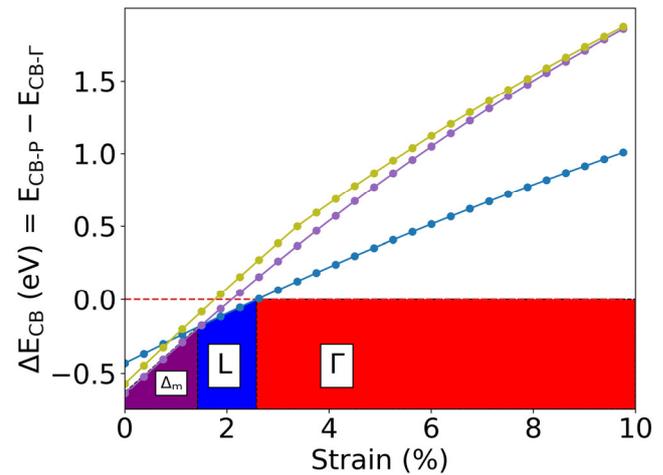

Figure 13: The difference between CB energies at the $\Gamma$ point and other k-points P ($\Delta E_{CB} = E_{CB-P} - E_{CB-\Gamma}$; with P = $\Gamma$, L, $\Delta_m$, and X) for isotropically strained GaP. Colored areas indicate at which k-point we found the CBM for the given value of tensile strain. The color scheme: $\Gamma$ (red), L (blue), $\Delta_m$ (purple), and X (olive).

For compressive strain, the CB at $\Gamma$ and L point increased in energy, while it decreased at the $\Delta_m$ and X points. As GaP is an indirect bandgap semiconductor at equilibrium, the nature of the bandgap thus did not change. For tensile strain, the CB energy at the $\Gamma$ and L points decreased strongly while we found a small increase at the $\Delta_m$ and X points. This led to a shift of the indirect bandgap from $\Delta_m$ to L at +1.43% tensile strain (figure 13). As the slope for the energy at the $\Gamma$ point was largest (figure 12), we found an indirect-to-direct (IDT) transition at +2.63% strain (figures 13 and S2).

Table 2: Change in the nature of bandgap for different III-V semiconductor materials for isotropic and biaxial strain.

| System | Transition | | | CBM transition path | |
|---|---|---|---|---|---|
| | Type[a] | Isotropic (%) | Biaxial (%) | Isotropic | Biaxial |
| Si | IDT | +10.31[b] | × | $\Delta_m \to L \to \Gamma$ | $\Delta_m \to K \to L$ |
| GaP | IDT | +2.63 | × | $\Delta_m \to L \to \Gamma$ | $\Delta_m \to L$ |
| GaAs | DIT | −1.56 | +3.52 | $\Gamma \to L \to \Delta_m \to X$ | $\Gamma \to \Delta_m$ |
| GaSb | DIT | −1.00 | +3.71 | $\Gamma \to L \to \Delta_m$ | $\Gamma \to \Delta_m$ |
| InP | DIT | −4.40 | +7.66 | $\Gamma \to X$ | $\Gamma \to \Delta_m$ |
| InAs | DIT | −7.41 | × | $\Gamma \to X$ | × |
| InSb | DIT | −5.18 | × | $\Gamma \to L \to \Delta_m$ | × |

[a] Direct to indirect (DIT) and indirect to direct (IDT) transitions.
[b] Estimated using linear extrapolation.
× No transitions within ±10% strain.

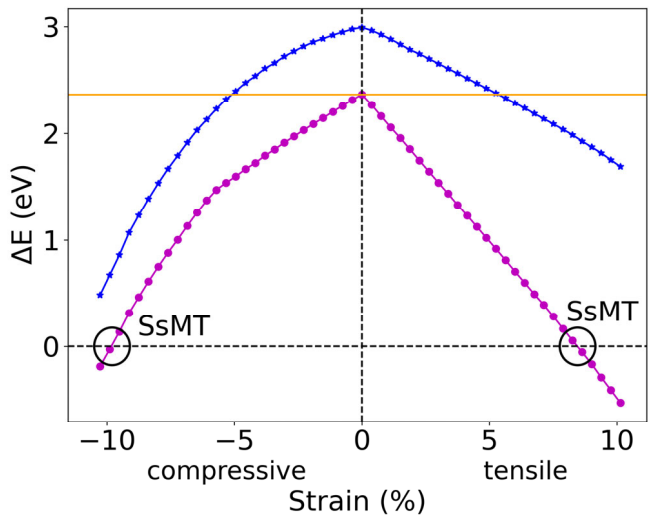

Figure 14: Biaxial strain effects on the bandgap of GaP. The energy difference between CB and VB at the Γ-point ($\Delta E_\Gamma$, blue), the bandgap ($E_g$, magenta), and the semiconductor to semimetal (SsMT) transitions are shown. The solid orange line indicates $E_g$ for the equilibrium structure (2.36 eV).

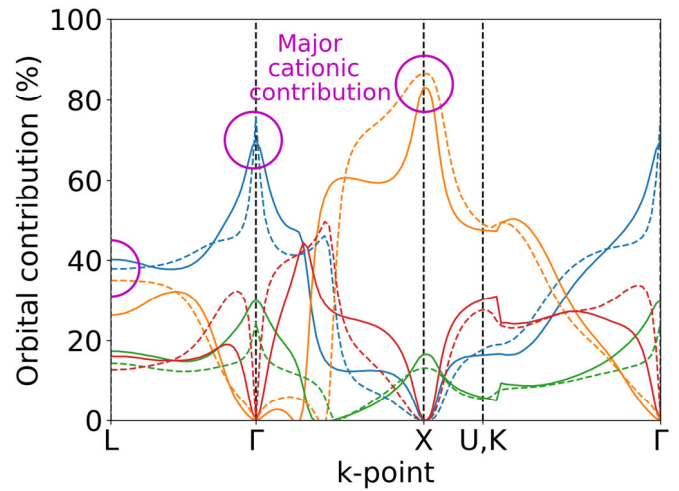

Figure 15: Atom resolved orbital contributions for GaAs and InAs CB. Solid sky blue: Ga(4s), dotted sky blue: In(5s), solid orange: Ga(4p), dotted orange: In(5p), solid green: As(4s) in GaAs, dotted green: As(4s) in InAs, solid red: As(4p) in GaAs, dotted red: As(4p) in InAs.

The result for biaxial strain is shown in figure 14 (band structures are shown in figures S1h,i). Here, we found no change in the nature of the bandgap throughout the entire range of compressive and tensile strain. The bandgap remained indirect throughout. For very high strain values, we found SsMTs at +8.45% and −9.83% strain.

The uniaxial conversion of the data in figure 14, as was explained for GaAs case, wouldn't have any further special interest (SI figure S3). Similar to GaAs, by combining biaxial and uniaxial strain in GaP one can in principle achieve IDT. However, this would require large uniaxial tensile strain (> 8 %), which can not be realized by thermal expansion (figure S4).

*4.3 Silicon, GaSb, InP, InAs, and InSb*

We also applied the approach outlined in detail for GaAs and GaP to other interesting semiconductor materials. Table 2 and 3 summarizes the main results for Si, GaSb, InP, InAs, and InSb. In all cases, the VBM stayed at the Γ point throughout the strain regimes applied. Thus, the position of the CBM in reciprocal space determined the nature of the band gap.

Table 3: Semiconductor to metal transition (SMT) and semiconductor to semimetal transition (SsMT) points for different III-V semiconductor materials under isotropic and biaxial strain. $\Delta E_\Gamma$ corresponds to the energy difference between CB and VB at the $\Gamma$ point.

| System | $\Delta E_\Gamma$ (eV) | SMT | | SsMT | |
|---|---|---|---|---|---|
| | | Isotropic (%) | Biaxial (%) | Isotropic (%) | Biaxial (%) |
| Si | 3.14 | +15.00[a] | × | × | +3.70, −6.50 |
| GaP | 2.99 | +13.00[a] | × | × | +8.45, −9.83 |
| GaAs | 1.81 | +6.67 | −7.86 | × | +8.00 |
| GaSb | 0.64 | +2.85 | −5.00 | × | +5.07 |
| InP | 1.43 | +8.20 | −9.90 | × | +10.38 |
| InAs | 0.36 | +2.10 | +4.74, −4.36 | × | × |
| InSb | 0.03 | +0.34 | +0.34, −0.34 | × | × |

[a]  Estimated using linear extrapolation.
×   No transitions within ±10% strain.

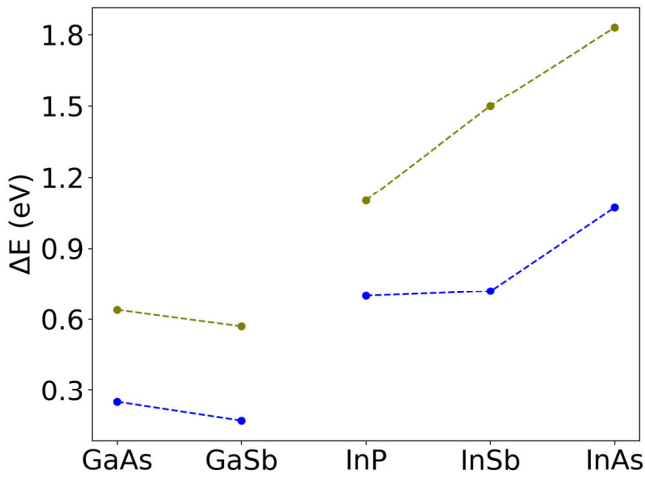

Figure 16: The relative energy differences in $\Gamma$–X (olive) and $\Gamma$–L (blue) at the CB for Ga and In series binaries.

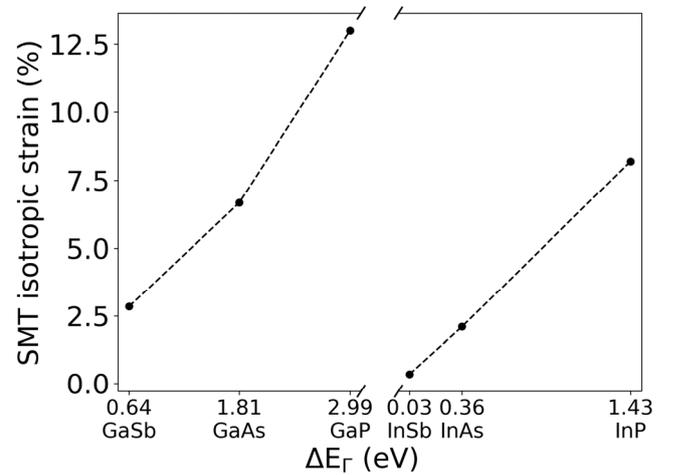

Figure 17: Correlation between SMT (isotropic strain) and the energy difference between CB and VB at the $\Gamma$-point ($\Delta E_\Gamma$) for Ga and In series binaries.

Si and GaP are indirect bandgap semiconductors in their equilibrium structure while the other materials discussed show direct bandgaps. Accordingly, Si and GaP showed IDTs while the other materials showed DITs. The strain values where these transitions were found, for isotropic and biaxial strain, are shown in columns 3 and 4 of table 2. For the isotropic strain case, the IDTs were found for tensile strain as already discussed for GaP in the previous section. The value for Si was so high (+10.31%) that it will certainly be out of range for any experiment. The DITs were found for isotropic compressive strain throughout, with strain values ranging from −1.00% (GaSb) to −7.41% (InAs).

For biaxial strain, DITs were found only for GaAs, GaSb, and InP. In all cases, a significant tensile strain would be necessary. For many materials, transitions were found to other k-points in reciprocal space where the nature of the bandgap stayed indirect. This is shown in the right-hand part of the table. [more details are shown in figures S5 and S6.]

Notably, In-based compound semiconductors showed the DIT points at much higher strain values compared to Ga-based materials. Figure 15 shows the contribution of the atomic orbitals to the CB for GaAs and InAs. At the decisive points in k-space ($\Gamma$, L, and X points) the group III elements showed the major contributions to CB, explaining the bandgap dependency on the group III element. Furthermore, the nature of the orbital contribution changes in k-space. While the Ga(4s) and In(5s) orbitals dominate at the $\Gamma$ point, the L point and the X point showed high contributions from the Ga(4p) and In(5p) orbitals. The energy gap between 5s and 5p orbital in In is much higher than 4s and 4p in Ga.

Table 4: The calculated equilibrium bandgaps, DIT points, and the DIT transitions compared with the experiments.

| System | Equilibrium bandgap (eV) | | Isotropic strain DIT (%) | | Transitions | |
|---|---|---|---|---|---|---|
| | Calculated | Experiment[a] | Calculated | Experiment | Calculated | Experiment |
| Si | 1.19 | 1.12 | — | — | — | — |
| GaP | 2.36 | 2.26 | — | — | — | — |
| GaAs | 1.47 | 1.42 | −1.56 | −2.10[b] | Γ→L | Γ→X[b] |
| GaSb | 0.64 | 0.73 | −1.00 | −0.70[b] | Γ→L | Γ→L[b] |
| InP | 1.43 | 1.34 | −4.40 | −3.51[c] | Γ→X | Γ→X[c] |
| InAs | 0.36 | 0.35 | −7.41 | −6.84[c] | Γ→X | Γ→X[c] |
| InSb | 0.03 | 0.17 | −5.18 | −3.09[c] | Γ→L | Γ→X[c] |

[a] Experimental bandgaps are at 300 K [38,129].
[b] Reference [28].
[c] Reference [28,157,158].
— For Si and GaP IDTs are in the tensile strain region. No experimental data available.

Therefore, changing group III from Ga to In increased the relative energy difference in Γ–X and Γ–L, figure 16. Under strain, the decrease of these relative energy differences ultimately resulted in the shift of CBM from the Γ to the L and/or X point (SI section S-VI). Therefore, the higher this relative energy difference, the higher the requirement of the amount of strain needed to reach the DIT point.

Table 3 summarizes the SMTs and SsMTs for the compound semiconductors investigated. These transitions depend on the closing of the CBM and VBM gaps. As the VBM always remains at the Γ point, these transition points therefore depend on $\Delta E_\Gamma$. Figure 17 shows the SMTs under isotropic strain for different systems in relation to their corresponding $\Delta E_\Gamma$. As the $\Delta E_\Gamma$ increased, so did the S(s)MTs.

In table 4 we compared our calculated results with the available experimental findings. The results match quite well. In experiments, the DIT points were measured in terms of applied hydrostatic pressure. Using the third-order Birch-Murnaghan equation [155] we converted the measurement in terms of strains. We used reference [156,157] for the bulk modulus and and its first derivative data for the conversion. For GaAs, the strain region when CBM was visible at the L point was very small (only 0.72 % strain window), figure 4. Therefore, we conclude that in the experiment this region was most likely missed (table 4, 3rd row last column). For InSb a deviation of 0.14 eV was found for the equilibrium bandgap. This, in turn, would result in the overestimation of the DIT point in our calculation (table 4).

## 5. Conclusions

We calculated the strain-induced bandgap variation for various III-V binary compounds focusing on GaAs and GaP for a detailed analysis. We investigated compressive and tensile strain in the range of ±10% around the equilibrium structure, which enabled the tuning of the bandgap over a wide range. Furthermore, we showed the presence of direct-to-indirect and indirect-to-direct transitions in the nature of bandgap of these materials based on the analyses of differences between valence and conduction band energies at the Γ-point ($\Delta E_\Gamma$) and the bandgap ($E_g$). Only 4 special k-points were found to be responsible for the direct-indirect transitions: Γ, L, $\Delta_m$, and X. The valence band maximum stayed at the Γ point throughout the strain regimes applied. Thus, the position of the conduction band minimum alone in reciprocal space determined the nature of the band gap. By combining the biaxial and uniaxial strain, we proposed a strategy for the realization of direct-indirect transitions in the regions where otherwise no transition could be achieved by single type of strain. With this work, we laid the foundation for further efforts with multinary compound semiconductors under strain.

## Data availability statement

All the density functional theory calculations data are available in the NOMAD repository (DOI: https://doi.org/10.17172/NOMAD/2022.08.20-2).

## Acknowledgments

We thank the German Research Foundation (DFG) for funding via the Research Training Group "Functionalization of Semiconductors" (GRK 1782). We acknowledge computational resources provided by HRZ Marburg, GOETHE-CSC Frankfurt, ZIH Dresden & HLR Stuttgart. We thank Prof. Kerstin Volz for discussions and continued support.

**ORCID iDs**

Badal Mondal   https://orcid.org/0000-0002-0522-1254
Ralf Tonner-Zech   https://orcid.org/0000-0002-6759-8559

# Supplementary Material

Systematic strain-induced bandgap tuning in binary III–V semiconductors from density functional theory

Badal Mondal[1,2], Ralf Tonner-Zech[1*]

[1]*Wilhelm-Ostwald-Institut für Physikalische und Theoretische Chemie,
Universität Leipzig, 04103 Leipzig, Germany*
[2]*Fachbereich Physik, Philipps-Universität Marburg, 35032 Marburg, Germany*

e-mail: ralf.tonner@uni-leipzig.de

## S I    Movies

The movies referenced here show the evolution of band structure in GaAs and GaP under different strain regimes. The band structures were calculated along the high symmetry path of zincblende structures. In all cases, the band energies were rescaled with respect to their corresponding VBM.

Figure S1a: GaAs under isotropic tensile strain
Figure S1b: GaAs under isotropic compressive strain
Figure S1c: GaAs under isotropic compressive strain zoomed in CB region
Figure S1d: GaAs under biaxial tensile strain
Figure S1e: GaAs under biaxial compressive strain
Figure S1f: GaP under isotropic tensile strain
Figure S1g: GaP under isotropic compressive strain
Figure S1h: GaP under biaxial tensile strain
Figure S1i : GaP under biaxial compressive strain

## S II    Indirect-to-direct (IDT) transition in GaP

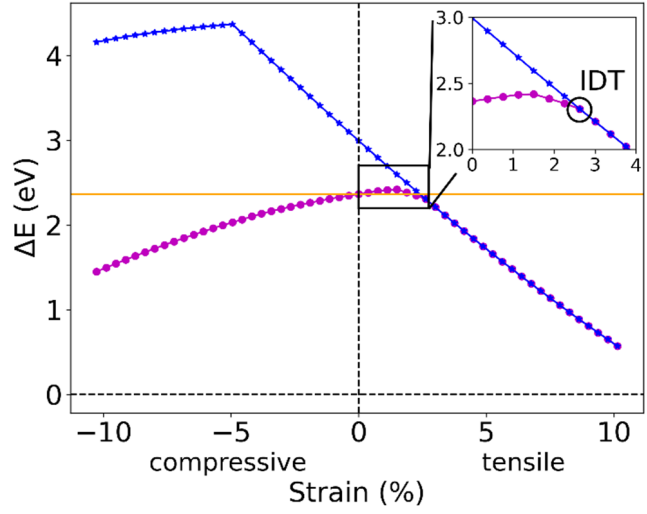

Figure S2: Isotropic strain effects on the bandgap of GaP. The energy difference between CB and VB at the Γ-point ($\Delta E_\Gamma$, blue), the bandgap ($E_g$, magenta), and the indirect to direct transition (IDT) are shown. The solid orange line indicates $E_g$ for the equilibrium structure (2.36 eV). The inset shows the region of the IDT with a finer grid of strain calculations resulting in higher resolution.

## S III  Uniaxial strain effect on bandgap of GaP

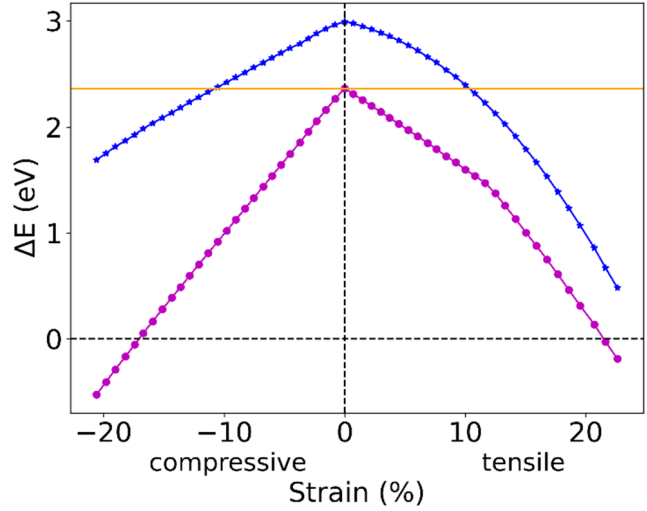

Figure S3: Uniaxial strain effects on the bandgap of GaP. The energy difference between CB and VB at the Γ-point ($\Delta E_\Gamma$, blue) and the bandgap ($E_g$, magenta) are shown. The solid orange line indicates $E_g$ for the equilibrium structure (2.36 eV).

## S IV   Indirect-to-direct (IDT) transition in GaP by combining biaxial and uniaxial strain

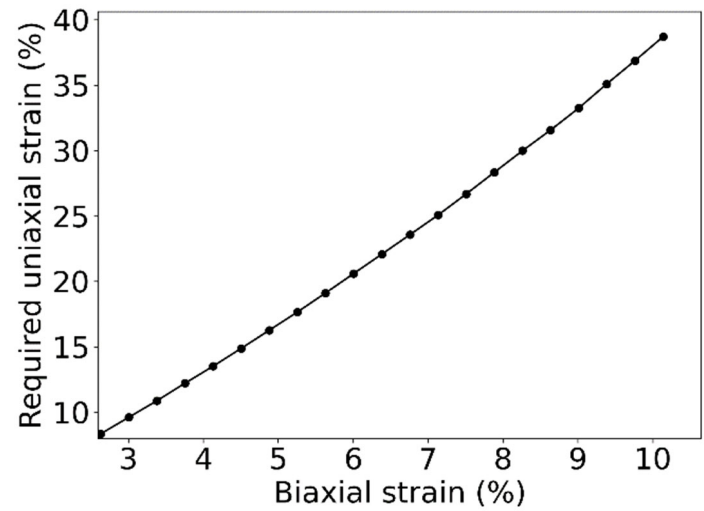

Figure S4: Variation of required uniaxial tensile strain for the indirect to direct transition (IDT) to take place in biaxial tensile strained GaP.

## S V   Bandgap variation with strain for III–V semiconductors

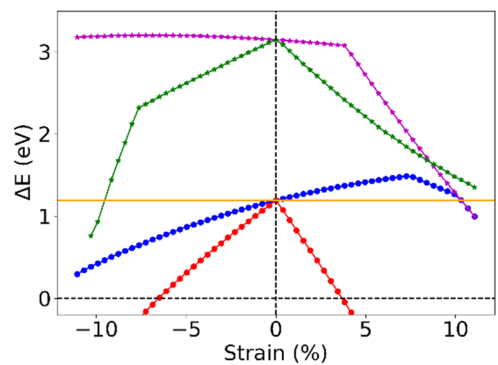

(a)  Si

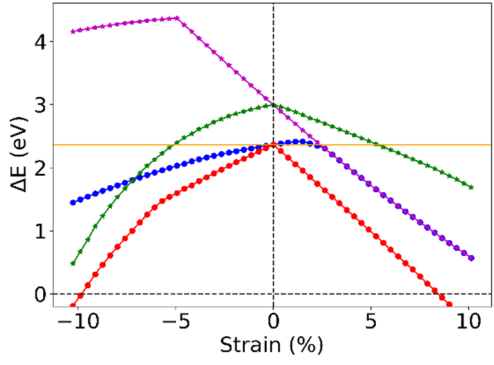
(b) GaP

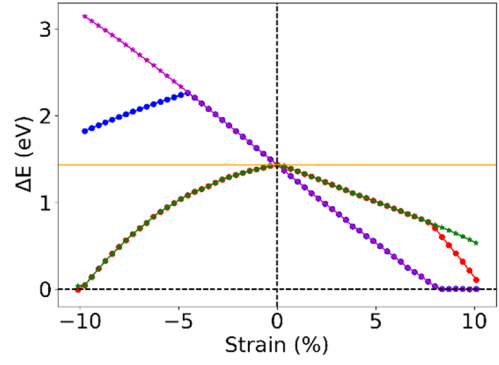
(e) InP

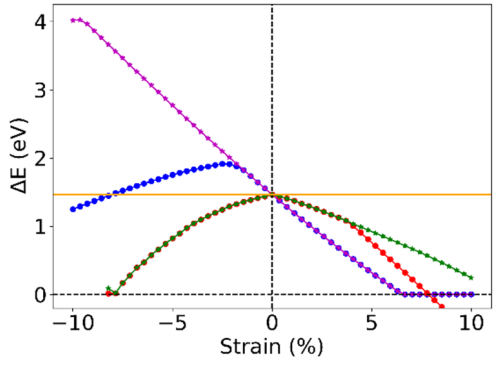
(c) GaAs

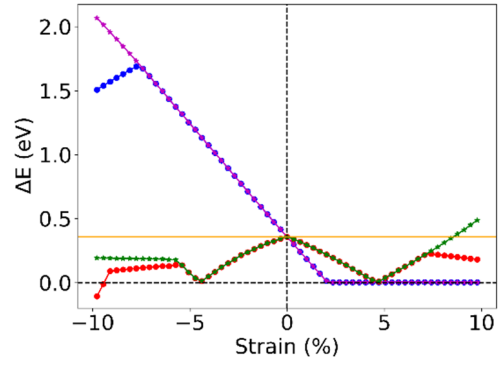
(f) InAs

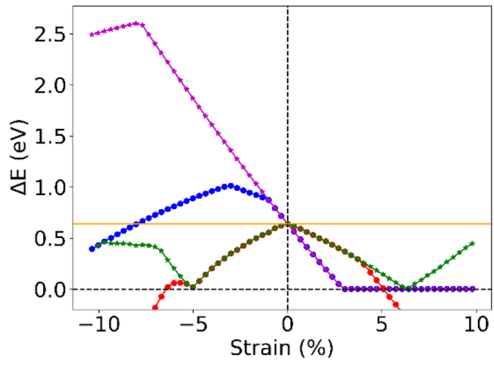
(d) GaSb

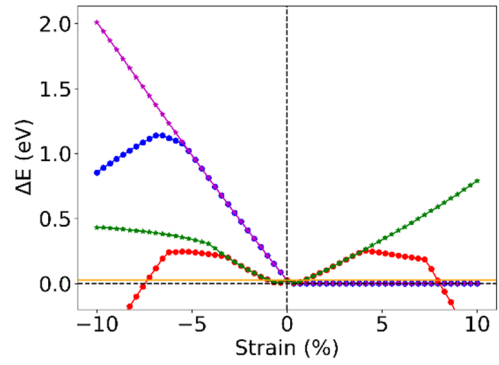
(g) InSb

Figure S5: Bandgap variation with strain for different III-V binary semiconductors. The energy difference between conduction and valence band at the $\Gamma$ point ($\Delta E(\Gamma)$, magenta for iso and green for bi) and the bandgap ($E_g$, blue for iso and red for bi) are shown as a function of biaxial (bi) and isotropic (iso) strain. The solid orange lines indicate $E_g$ for the equilibrium structures.

# S VI  CBM transition path for different III–V semiconductors

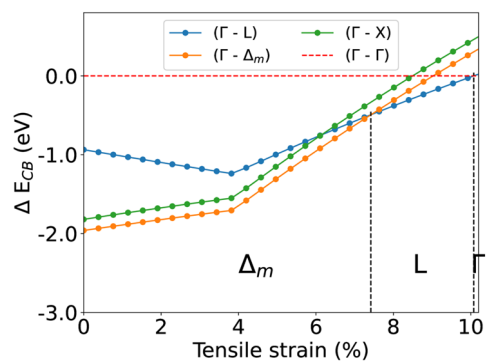

(a) Si

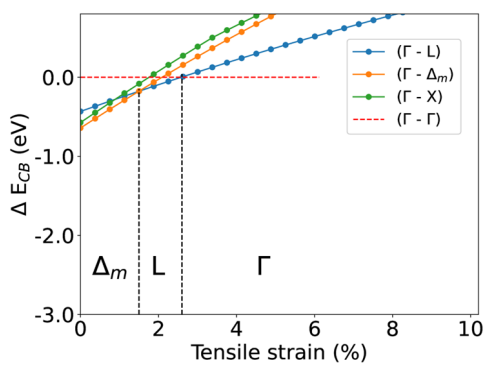

(b) GaP

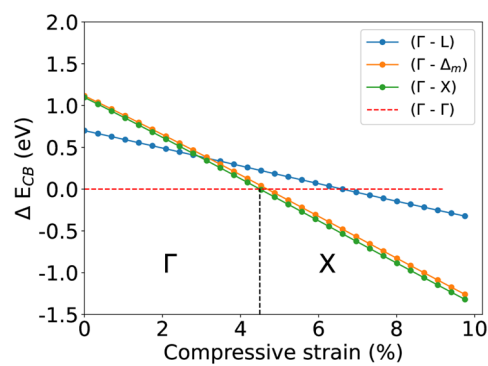

(e) InP

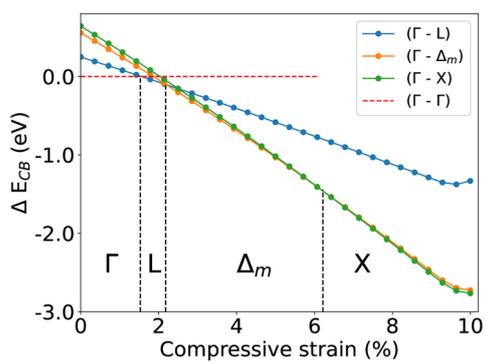

(c) GaAs

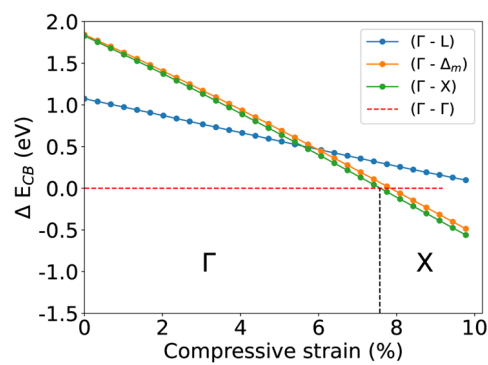

(f) InAs

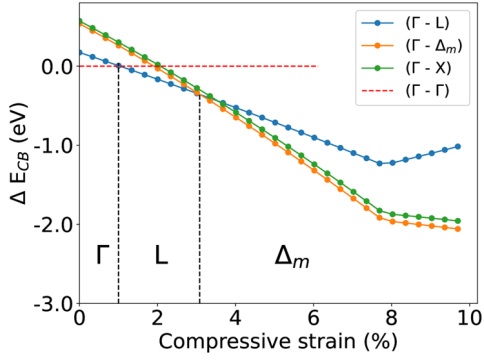
(d) GaSb

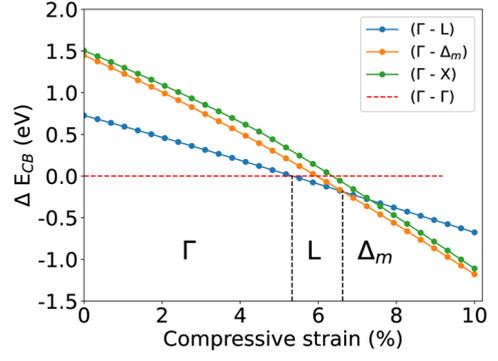
(g) InSb

Figure S6: The difference between conduction energies at the $\Gamma$ point and other k points P ($\Delta E_{CB} = E_{CB\text{-}P} - E_{CB\text{-}\Gamma}$ ; with P = L, $\Delta_m$, and X) for isotropically strained III-V binary semiconductors. The enclosed areas between the lines and the x-axis indicate at which k-point we find the conduction band minima for the given values of strains.

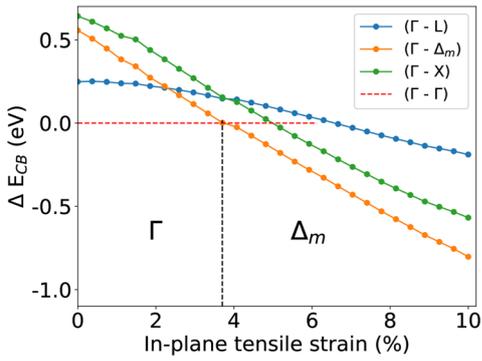
(a) GaAs

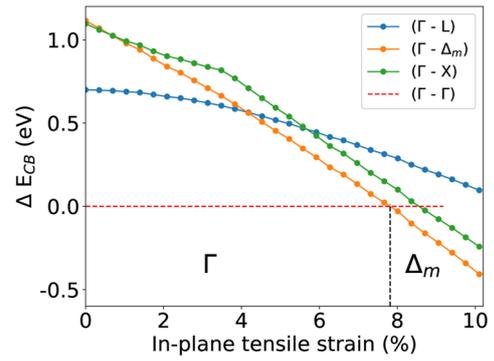
(c) InP

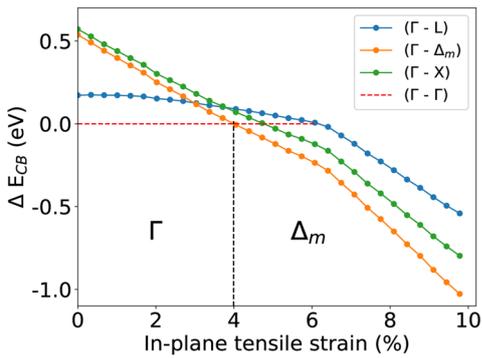
(b) GaSb

Figure S7: The difference between conduction energies at the $\Gamma$ point and other k points P ($\Delta E_{CB} = E_{CB\text{-}P} - E_{CB\text{-}\Gamma}$ ; with P = $\Gamma$, L, $\Delta_m$, and X) for biaxially strained III-V binary semiconductors. The enclosed areas between the lines and the x-axis indicate at which k-point we find the conduction band minima for the given values of strains.